\documentclass
[reprint,
]{revtex4-2}

\usepackage{graphicx}
\usepackage{dcolumn}
\usepackage{bm}

\begin{document}

\preprint{APS/123-QED}

\title{Delta-isobar resonance effects studied by $\tau\sigma$ summed strengths and \\ 
 nuclear matrix elements for $\beta$ and $\beta\beta$ decays}  

\author{Hiroyasu Ejiri}
\email{ejiri@rcnp.osaka-u.ac.jp}
 \affiliation{Research Center for Nuclear Physics, Osaka University, Osaka 567-0047, Japan
 }%

\date{\today}
\begin{abstract}
Nuclear matrix elements (NMEs) for neutrinoless $\beta\beta$ decays (DBD)
and inverse-$\beta$ decays are crucial for studying $\nu$ properties beyond the standard model and astro-$\nu$ nuclear interactions.  The NMEs consist mainly of  the  isospin ($\tau$) spin ($\sigma$) component, and the $\tau\sigma$ strength (square of the $\tau\sigma$ NME) is studied experimentally by charge exchange reactions (CERs). 
 The summed Gammow-Teller (GT; $\tau\sigma$) and spin dipole (SD; $\tau\sigma Y_1$) strengths measured by CERs are shown to be reduced half with respect to the nucleon-based sum rule limits.  The  $\tau\sigma$ NMEs are shown to be quenched due to the non-nucleonic $\Delta$-isobar resonance effect on the basis of the measured strengths and the QRPA  analysis with effective NN and N$\Delta$ $\tau\sigma$ interactions. The quenching effect is  incorporated  by using an effective $\tau\sigma$ coupling of $g_{\tau\sigma}^{\Delta}/g_{\tau/\sigma}$ $\approx$0.7 with $g_{\tau\sigma }$ being the coupling for a free nucleon.  The quenching effect is applied to the $\tau\sigma$ components of the weak, electromagnetic and nuclear interaction NMEs. Impact of the $\Delta$-resonance effect  on $\nu$ studies in nuclei is discussed.\\

             
\end{abstract}

\maketitle



\section{ Introduction} 
Neutrino ($\nu$) properties such as the Majorana nature, the $\nu$-mass, and the right-handed weak currents beyond the standard model are studied by neutrinoless double beta decays ($\beta\beta$) as discussed in \cite{doi85,avi08,ver12,ago23}.
Astro-$\nu$ productions, syntheses and oscillations are studied by $\nu$-nuclear reactions (inverse-$\beta$ decays).  Nuclear matrix elements (NMEs) for nuclear $\beta\beta$ and inverse-$\beta$ decays are crucial for studying these $\nu$ properties of the current astro-particle physics interests. 

 The NMEs, reflecting complex nuclear structures, are very sensitive to all kinds of nucleonic and non-nucleonic correlations. Major components of the NME are the isospin ($\tau$) axial-vector (spin $\sigma$)  ones. Accordingly,  the $\beta\beta$ and inverse-$\beta$ NMEs depend much on nucleonic and non-nucleonic $\tau\sigma$ correlations, which are induced by various $\tau\sigma$-type  nuclear interactions.  The $\beta$ NMEs have been discussed extensively  in \cite{eji68,eji78,eji82,eji83,tow87,eji00,eji14,eji15,eji19,gys19,cor24} and references therein, and the $\beta\beta$ NMEs in \cite{eji00,eji19,suh98,fae98,men11,suh12,bar13,suh17,eng17,jok18,sim18,eji22} and references therein. 
\begin{figure}
\hspace{0cm}
\vspace{-0.3cm}
\includegraphics[width=0.5\textwidth]{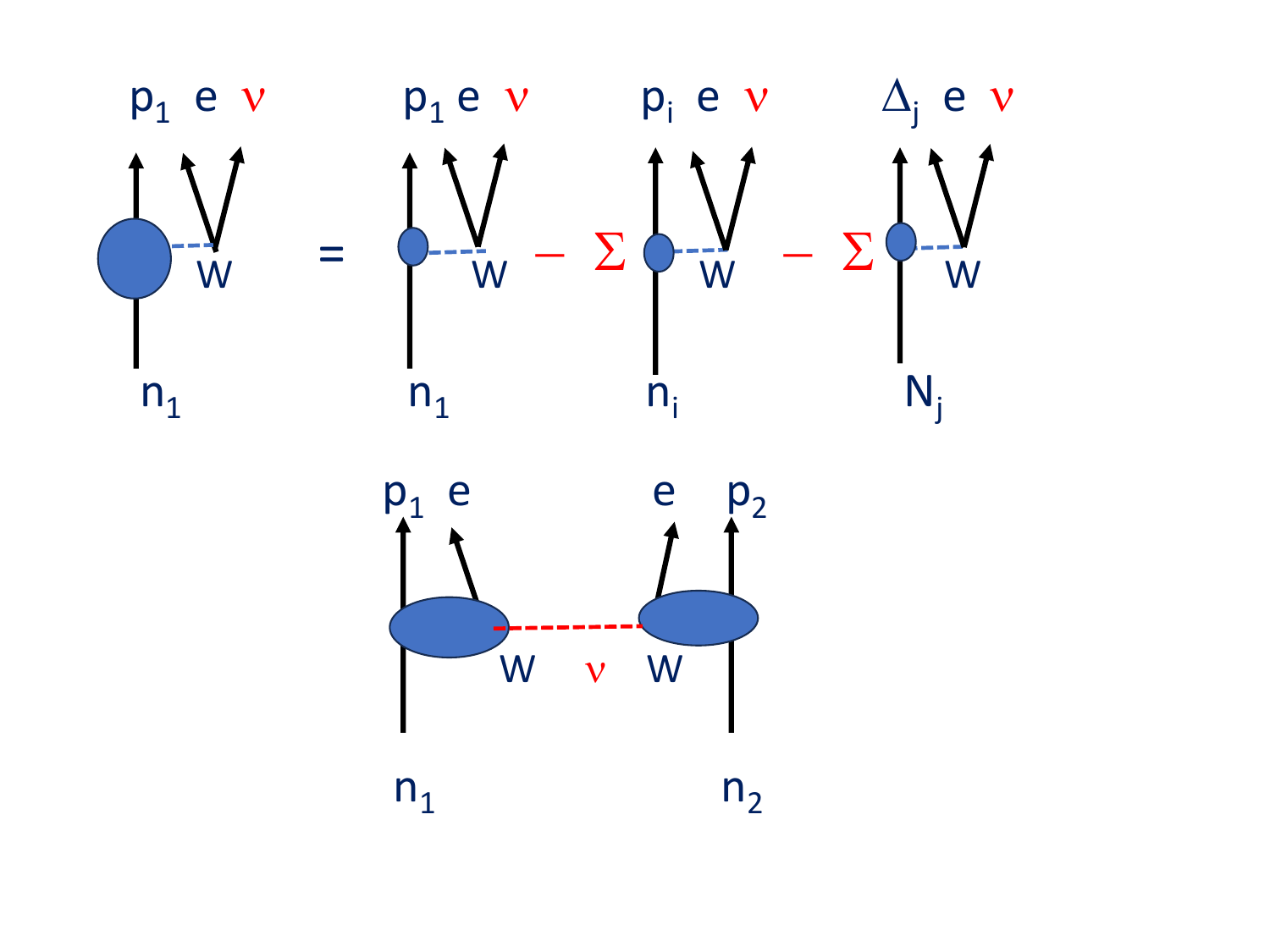}
\vspace{-1cm}
\caption{Schematic diagram of a QP $\beta$ decay (top) and that of a  $\beta\beta$ decay (bottom). $\nu$ : Majorana neutrino. The NME is given schematically by a coherent sum of a QP NME for n$_1\rightarrow$p$_1$ and a GR$_{\rm N}$ NME for  
$\Sigma_i$ [n$_i\rightarrow$p$_i$] and 
 a GR$_{\Delta}$ NME for $\Sigma_k[$ N$_k\rightarrow \Delta_k$].  The GR$_{\rm N}$ and the GR$_{\Delta}$ are mixed with the QP by the NN and N$\Delta$ interactions via meson-exchange interactions. 
\label{figure:fig1}} 
\vspace{-0.5cm}
\end{figure}

A nucleus is composed mainly by nucleons (N) with $\tau,\sigma$=1/2,1/2, and thus the NN $\tau\sigma$ correlations are the correlations to be considered in $\beta\beta$ and inverse-$\beta$ NMEs. The delta isobar ($\Delta=\Delta_{33}$) with $\tau,\sigma$=3/2,3/2 is strongly excited by the quark $\tau\sigma$ flip from a nucleon, and the N$\Delta$ $\tau\sigma$ correlations are the key non-nucleonic  correlations to be considered as well for the $\beta\beta$ and inverse-$\beta$ NMEs.  
The NMEs  discussed in the present work are such $\tau\sigma$ ones that couple directly with the $\Delta$. They are involved in axial-vector weak, isovector magnetic and $\tau\sigma$ nuclear interactions.

The strong repulsive ($V>$0) NN $\tau\sigma$ interaction pushes up the nuclear $\tau\sigma$ strength  (square of the $\tau\sigma$ NME) to form the $\tau\sigma$ NN giant resonance (GR$_{\rm N}$) in the 10-15 MeV region, leaving a little $\tau\sigma$ strength in low-lying quasi-particle (QP) states. Then $\tau\sigma$ NMEs for low-lying states are much reduced due to this $\tau\sigma$ NN correlation.    

Likewise,  the  strong repulsive N$\Delta$ $\tau\sigma$ interaction pushes up the $\tau\sigma$ strength to form the N$\Delta$ giant resonance (GR$_{\Delta}$) at the high excitation region around the $\Delta$ mass ($\approx$ 300 MeV), and thereby  shifts  some $\tau\sigma$ strength from the N region around 0-30 MeV to the $\Delta$ region around 300 MeV, resulting in considerable reduction of the summed $\tau\sigma$ strengths in the N region and thus in severe quenching of the axial-vector components of the $\beta\beta$ and inverse-$\beta$ NMEs in the N region. 

The present report is concerned mainly with the gross non-nucleonic effect associated with the GR$_{\Delta}$ . This effect is studied experimentally by investigating the reduction of the summed $\tau\sigma$ strength in the N region. 
Actually, there are many other nucleonic correlations that affect more or less individual nuclear states and their NMEs as discussed extensively by using various kinds of nuclear models  in the review \cite {eji19}, but they are not discussed in the present work.

 The $\tau\sigma$ strengths in DBD nuclei have been well studied experimentally in the wide nucleonic excitation region (N region) of $E\approx$0 - 30 MeV by using charge exchange reactions (CERs). Then, the non-nucleonic reduction effect is well studied experimentally by investigation the summed $\tau\sigma$ strengths there.
 
The present work aims to study for the first time the non-nucleonic GR$_{\Delta}$ effects on $\tau\sigma$ (axial-vector) NMEs for medium-heavy DBD nuclei on the basis of the experimental summed  $\tau\sigma$ strengths and GR energies measured by the medium-energy CERs. The gross feature of the non-nucleonic quenching effect is discussed  by using the  quasi-particle random phase approximation (QRPA(N$\Delta$)) with both the NN $\tau\sigma$ and N$\Delta$ $\tau\sigma$ nuclear interactions.  The DBD nuclei discussed are $^{76}$Ge, $^{82}$Se, $^{96}$Zr, $^{100}$Mo, $^{130}$Te, $^{136}$Xe, and $^{150}$Nd, which are of current interest.  They are the nuclei with $N$(neutron number)$\gg Z$(proton number) since $\beta$/EC (electron capture) decays are forbidden. 

The QP $\beta$ and $\beta\beta$ transition diagrams  are schematically shown in Fig. 1. Here
GR$_{\rm N}$ is given by a coherent sum of $j$-neutron (n$_{j}$ )  to $j$-proton (p$_{j}$) excitations, while GR$_{\Delta}$ by  a coherent sum of N$_k$ to $\Delta_k$ excitations. 
The $\tau\sigma$ NME for a low-lying QP state is reduced from the QP NME for n$_1$ to p$_1$ by the negative-phase admixtures of the NMEs for the GR$_{\rm N}$ and GR$_{\Delta}$. Here the admixture is due to the repulsive NN and N$\Delta$ interactions. 

Actually,  the $\Delta$ effect on $\tau\sigma$  NMEs involved in weak, electromagnetic and nuclear (strong) interactions has been discussed before  \cite{ose79,boh81,eji82,sag82,eji83,ost85,tow87,ris89,ost92,kir99,cat02}. These works suggest severe reduction (quenching) due to the $\Delta$ effect.  Exchange (meson) and 2-body (2B) currents are associated with the $\Delta$ via the meson-exchange between 2 nucleons (2B), and their effects on $\beta$ and $\beta\beta$ NMEs are discussed very well theoretically \cite{gys19,men11,cor24}.  

Axial-vector ($\tau\sigma$) $\beta$ and $\beta\beta$ NMEs are discussed as given in reviews and
 references therein \cite{eji00,eji19} and also in recent works  \cite{eji22,eji14,eji15,eji19a}. There the experimental GT (Gamov-Teller) and SD (spin dipole) NMEs are shown to be reduced much more than expected from nucleonic $\tau\sigma$ correlations, and some reduction effects  due to non-nucleonic correlations such as the N$\Delta$ are discussed.

\section{ Summed  GT strengths }
\subsection { Summed GT strengths in the N region}
We first discuss summed GT $\tau^{\pm}$ strengths defined as $S^{\pm}=\Sigma B({\rm GT}^{\pm})$ with $B({\rm GT}^{\pm}) \equiv |M({\rm GT}^{\pm})|^2$ for 0$^+\rightarrow 1^+$ transition in the N region. The Ikeda-Fujii-Fujita sum-rule  \cite{ike63} is given as $S_{\rm N}({\rm GT})=S^--S^+=6T_{\rm z}=3(N-Z)$.  This rule is based on nuclear models with N (nucleon: i,e $A-Z$ neutrons with $t_z=1/2$ and $Z$ protons with $t_z$= -1/2) and NN correlations, but no non-nucleonic baryons like $\Delta$ and their correlations.

 The non-nucleonic $\Delta$ effect on the $\tau\sigma$ NMEs is studied by investigating if the sum in the N region  gets smaller than the sum rule by shifting the strength up from the nucleonic N region around 0-30 MeV to the non-nucleonic $\Delta$ region around 300 MeV.  Recently  we have measured the $B({\rm GT^-})$ for low-lying QP states relevant to 2$\nu\beta\beta$ NMEs within the standard model at RCNP Osaka by using the  ($^3$He,$t$) CERs
 \cite{aki97,thi12,thi12b,pup11,gue11,thi12a,pup12,fre13,eji16a,zeg07,dou20} on medium-heavy DBD nuclei  and others. We reanalyze the data to extract  the GT and SD summed strengths and their GR energies relevant to the 0$\nu\beta\beta$ NMEs beyond SM and astro-$\nu$ NMEs. Note that the GT and SD NMEs studied by CERs with the strong interaction are the same GT and SD NMEs for weak and electromagnetic interactions, although their interaction couplings are very different.

The unique features of the present CER are as follows \cite{eji00,eji19}. i. The medium-energy ( $E\approx$ 0.42-0.45 GeV) projectile excites  preferentially GT(1$^+$) and SD(2$^-$) states because of the dominant $\tau\sigma$ (axial-vector) nuclear (strong) interaction, namely $V_{\tau\sigma} \gg V_{\tau\tau}$, $V_{\rm T}$ where $V_{\tau\sigma}$, $V_{\tau\tau}$ and $V_{\rm T}$ are the axial-vector,  the vector and the tensor interactions \cite{eji00,eji19,fra85}. The distortion potential of $V_0$ gets minimum at the medium energy $E$.
 ii. The high energy-resolution (10$^{-5}$) spectrometers  for the incident and emitted particles make it possible to measure well the GT and SD strengths, being free from backgrounds.  The distortion effect is well evaluated experimentally. Thus one can get experimentally reliable GT and SD $\tau\sigma$ NMEs.  In fact, GT NMEs derived from the CER experiments, after adequate corrections for the $V_{\rm T}$  effects,
agree within a few $\%$ with GT NMEs derived from $\beta$-decay and EC experiments  \cite{eji00,eji19, fre13,ell23}.  iii. The present CERs with the nuclear probe is quite feasible, while the CERs with weak probes like neutrinos and muons are not realistic experimentally because of the extremely small cross section, the very low beam intensity and the poor energy resolutions by many orders of magnitude than those of the present CER experiment.
 
 GT$^-$ and SD$^-$ strengths at low lying QP states are shifted to the GR$_{\rm N}$(GT) and GR$_{\rm N}$(SD) and to the possible GR$_{\Delta}$(GT) and GR$_{\Delta}$(SD), as shown in Fig. 2. The isobaric analogue state (F in Fig. 2) is the Fermi ($\tau^-$)-type GR$_{\rm N}$(F). 
\begin{figure}
\hspace{-0.8cm}
\vspace{-0cm}
\includegraphics[width=0.45\textwidth]{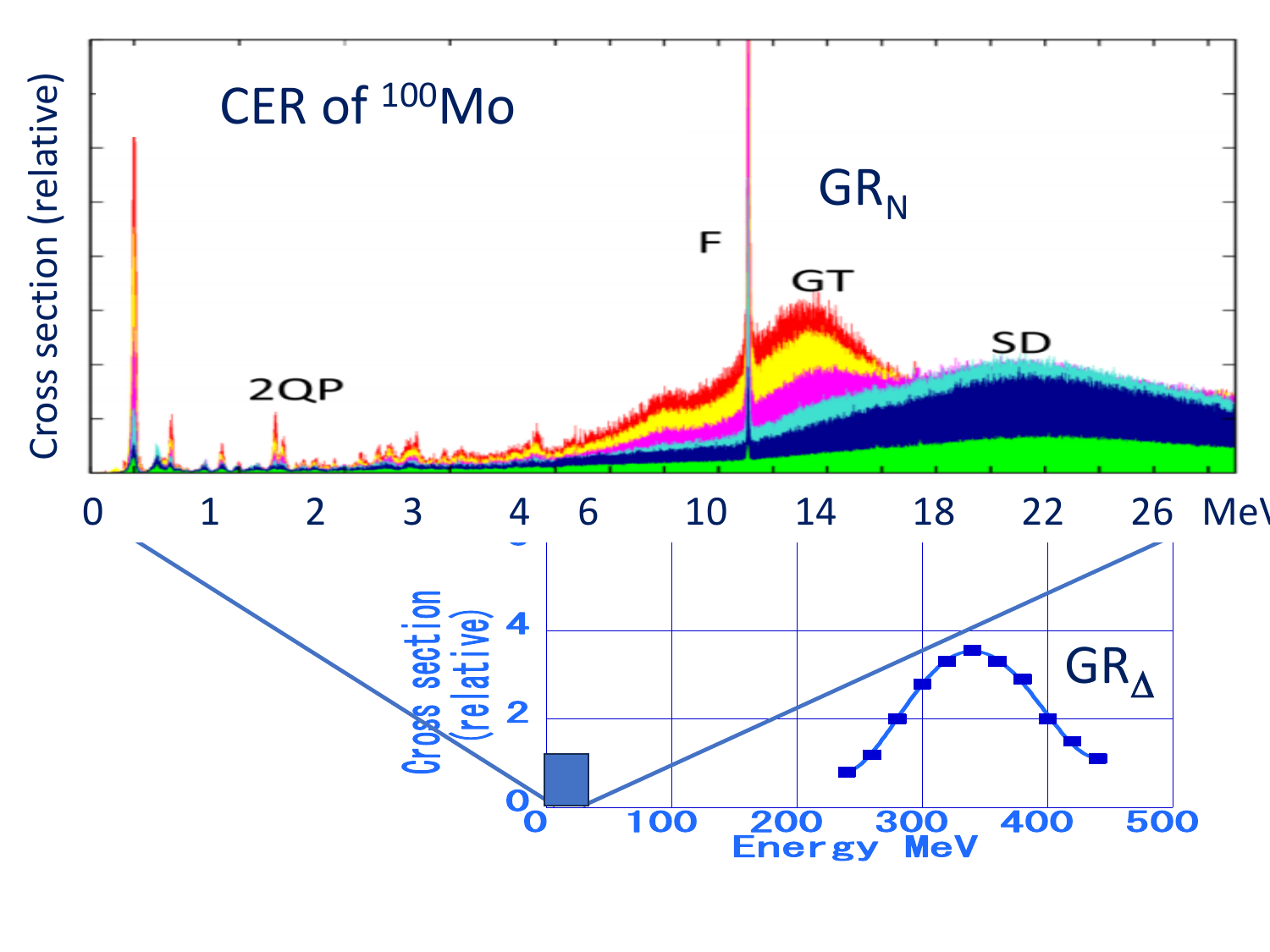}
\vspace{-0.5cm}
\caption{The ($^3$He,$t$) energy spectrum (insert: enlarged), together with the possible
 GR$_{\Delta}$. F, GT, and SD are GR$_{\rm N}$(F),
GR$_{\rm N}$(GT), GR$_{\rm N}$(SD). Red, yellow, pink, light blue, blue and green spectra are for angle intervals of 0-0.5, 0.5-1, 1-1.5, 1.5-2, 2-2.5, and 2-2.5, each in deg. The $l$=0 GT and the $l$=1 SD components dominate at 0-0.5 deg. and 1.5-2 deg.  \cite{eji22,thi12a}.
\label{figure:fig1A}} 
\vspace{-0.4cm}
\end{figure}

The GT strength from the 0$^+$ state ($|0>$) to the GT ($J_{\rm f}$=1$^+$) state is expressed by 
\begin{equation}
B({\rm GT})=|<f||\sigma||0>|^2, ~\delta n=\delta l=0, ~\delta j=0,1,
\end{equation}
with $n,l,j$ being the radial-node number, the angular-node number, and the angular momentum. The GT $\tau ^- (\beta^-)$ transitions with the transition operators of $\tau$ and $\sigma$ are limited, in thems of nuclear shell models, to the neutron$\rightarrow$proton transitions between the same $n,l$ shell-orbit in the excitation region mainly below 25 MeV (i.e. no jump in energy to higher $n$ shells in the region beyond $E\approx$ 20 MeV) with no changes of  the radial and angular nodes of $n$ and $l$.  Thus, the GT transition is allowed only within the same $N$$\hbar \omega$ shell, and not to higher $N'\hbar \omega$ shells  in the higher
 excitation continuum region. The emitted $t$  is limited to the forward direction of $\theta\approx$0 deg. because of $\delta l$=0. Thus the GT excitation is limited mainly in the low excitation region around 0 -20 MeV because of $\delta n$=0 and at the low (forward) angle region around $\theta$=0 deg. because of $\Delta l$=0.
The selection of the $\delta n$=0 component is crucial for identification of GT in the high excitation region as remarked in the section 3.7 of \cite {eji19}. 

We first discuss the GT($\delta l=0, \delta n$=0) strength in the CER energy spectrum at 0 deg. in the 0-28 MeV region. It consists mainly of the $\delta l$=0 component with the small admixtures of $\delta l$=1, 2 ones, which are corrected for in the analysis.  The $l$=0 component is due to the low-lying QP GT states, the GR$_{\rm N}$(GT) around 10-15 MeV and the quasi-free scattering (QFS) in the 10-28 MeV continuum region \cite{aki97,dou20,pha95,gue11}.  The GR$_{\rm N}$(GT) width of $\Gamma$(GT)$\approx$ 8 MeV reflects the large spreading of the GR$_{\rm N}$(GT) strength to 2-particle 2-hole states \cite{ber82,gam20}. The QFS consists of the QF-GT
 component, including the GT strength spreading into the 2p-2h and the upper-isospin GT strength and the QF non-GT ($\delta n\ne 0$)  strength. 

Then the summed GT strength $S^-$(GT) is the sum of the $B(\rm GT)$ values for the low-lying QP GT states, the GR$_{\rm N}$(GT) and the QF-GT continuum. The non-GT ($\delta n\neq$0) strength is around 7$\%$ of the $S^-$(GT), being consistent with the low-energy part of the calculated non-GT $J^{\pi}$=$1^+$ strength  (iso-vector monopole and other $\Delta n$=1) around 25-36 MeV. 
The summed  strengths of $S^-$(GT)  for  DBD and Sn nuclei are around 47$\pm{5}\%$ of $S_{\rm N}$(GT) \cite{ike63} as shown in Fig. 3. The $l$=0 strength extending beyond 25 MeV is mainly non-GT $(\delta n$=1,2) excitations. The individual CER cross-sections at 0 deg. are affected a little by the tensor interaction contribution,  but the effects cancel out more or less in the summed strength.  

The $\tau ^+$ strength in the DBD nuclei has been investigated by (d,$^2$He) and ($t,^3$He) reactions 
 \cite{gue11,doh08,gre08}. It is around 5$\%$ of $S_{\rm N}$(GT) because of the blocking of the p to n transition by the large excess neutrons in the $\beta^-\beta^-$-DBD nuclei. The non-GT ($\delta n$$\neq$0) $\beta^+$ strength extends beyond 10 MeV in the QFS region, being not blocked by the excess neutrons \cite{ray90,con92}.
Then one gets $S^--S^+\approx 0.42\pm$ 0.05 of $S_{\rm N}$(GT)  for the summed GT ($\delta l=\delta n=0$) strengths up to 28 MeV.  The error includes systematic errors in corrections for the small non-GT ( i,e. $\delta l \ne 0, \delta n\ne 0$) components. The $\delta l$=0 strength extends due to the 2p-2h spreading certainly beyond 28 MeV, but is mainly of  non-GT with $\delta n\ge1$. The GT component beyond 28 MeV is evaluated to be an order of 5 $\%$ of  $S_{\rm N}$(GT). Then, including this, we get $S^--S^+\approx 0.47\pm$ 0.07 of $S_{\rm N}$(GT). This reduction is nearly same with the reduction around 0.5 in other nuclei \cite{gaa81,gaa83,eji00, eji19}.

\subsection{ Comments on (p,n) CERs reactions}
GT and SD strengths for non-DBD nuclei have been extensively studied in the 1980' and 1990' by using medium energy (p,n) and (n,p) reactions as given in the review \cite{eji00,eji19} and refs therein. They excite preferentially GT and other $\tau\sigma$ states/resonances as the present ($^3$He,$t$) reactions because of the dominant $V_{\tau\sigma}$ interaction. The summed strengths of $S_{\rm N}$(GT) in the N region of around $E$=0-30  MeV are found to be around 50 $\%$ of the sum rule limit of 3$(N-Z)$ in a wide mass region of the mass number $A$=30 - 208, including $^{90}$Zr \cite{gaa81,gaa83}. These are simiar to the present ($^3$He,$t$) CER results for DBD nuclei.

The (p,n) reaction on $^{90}$Zr shows the large summed GT strength of  $S^--S^+\approx 3(N-Z)$ below 50 MeV \cite{wak97}. It is claimed that the 90$\%$ of the sum rule limit \cite{ike63} is the minimum value, and suggests a large strength of the order of 10-20 $\%$ of the limit in the higher region of 50 - 60 MeV.  A similar work on $^{90}$Zr reports 88$\pm6$$\%$ of the sum rule \cite {yak03}. These works are in support of the sum rule, but are contrasted to the other CER data for the same $^{90}$Zr and others \cite{gaa81,gaa83}.

 The (p,n) spectra for $^{90}$Zr in both \cite {gaa81,gaa83} and \cite {wak97} show the same strong GT ($\delta n$=0) GR$_{\rm N}$ around $E\approx$ 10 MeV of the $\delta n=0$ region, extending up to around 16 MeV and the similar continuum spectrum increasing slowly from 16 MeV to 50 MeV and beyond. This high-excitation region corresponds just to the non-GT $\delta n$=1, 2, 3 region. The strength in this continuum is not included into the summed GT strength, resulting in the severe quenching with respect to the sum rule limit in \cite{gaa83} and other (p,n) works. On the other hand,  the strength in this continuum, except the IVSM (Isovector spin monopole) contribution, is considered to be included in the summed GT strength, resulting in the full sum-rule limit in \cite{wak97}.  
The observed strength in the $\delta n\ge$1 continuum region beyond 25 MeV should be corrected for the main non-GT components with $\delta n$=1-3.

 Actually the observed strength around 30 MeV is consistent with the calculated $\delta n$=1 strength.  Then the large observed GT strength of $S_{\rm N}$(GT)$\ge$90 $\%$ of the sum rule value could be reduced to about 50-60 $\%$ of it if the large non-GT strength with $\delta n\ge$1 in the high excitation region would be corrected for.   The systematic studies of the (p,n) CERs in the mass region of $A$=13-90 show some 50-60 $\%$ of the sum rule imit \cite {goo84}.   The  (p,n) reaction on the much lighter nucleus of $^{48}$Ca gives the $S^--S^+$=52 $\%$ of $3(N-Z)$ below 30 MeV \cite{yak09}. The measured $S^-$ up to 30 MeV is certainly smaller  than  the 2p-2h calculation without the $\Delta$ \cite{gam20}. 

The RPA calculation for the $^{90}$Zr shows that the strength in the GR$_{\rm N}$ $\delta n$=0 region below 25 MeV is GT with $\delta n$=0, but the large strength in the $\delta n\ge1$ region beyond 30 MeV is not GT \cite{ost85}. On the other hand, the calculation including coupling with 2p-2h shows a large spread of the GT strength of around 40 $\%$ into the higher excitation region of 25 - 40 MeV in case of no coupling with $\Delta$ \cite{ber82}.  

In the (p,n) experiment, a wide excitation energy region is covered by measuring emitted neutrons by TOF (time of flight) method, while in the present ($^3$He,$t$) experiment the emitted tritons are measured by the spectrometer at RCNP with the limited region around 0-30 MeV in the excitation energy. Then we have corrected for the non-GT ($\delta \ne$ 0) contribution in the continuum region of 18-28 MeV, and then evaluates the possible small GT contribution beyond 28 MeV to be included in the summed GT strength.   

The CER strength in the continuum region beyond 20 MeV is also treated as QFS (quasi-free  scattering), where a nucleon in the target nucleus is excited to the unbound continuum region.  It includes a little GT strength in the lower excitation region below 20 MeV, but is mainly non-GT strengths with $\delta n\ge$1, and thus all the QFS at the higher excitation is not included into the GT with $\delta n$=0, being in contrast to the (p,n) work in \cite {wak97}.

It is noted here that the present GR$_{\Delta}$ is based on the experimental data showing the severe quenching of the summed strengths with respect to the sum rule limit \cite {ike63}. The recent theoretical work on 2B \cite {gys19} shows also the severe quenching of the summed GT strength.  In fact, the present  GR$_{\Delta}$ and other $\Delta$ effects and the exchange (2B) effects are based mainly on the non-nucleonic quenching effect of $\Delta$ located far above the N region. Thus they would be not be right if all the GT strengths would be found experimentally or theoretically in the N region of 0-50 MeV as claimed in \cite {wak97}. So further theoretical and experimental studies of the summed GT strengths are interesting.

\subsection{ Summed SD strength in the N region}
The SD strength  for the $\delta n$=0,1 and $\delta l=1$ $\tau\sigma$ transitions is given by  $B(\rm SD)=|<f||r[\tau\sigma Y_1]_J||0>|^2$ with $J=0,1,2$. The GR$_{\rm N}$(SD) is located about 1 $\hbar \omega \approx$ 10-8 MeV above the GR$_{\rm N}$(GT) as shown in Fig. 2. The large width around 15 MeV is due to the spreading of the SD strengths to 2p-2h states and the $J$ dependence of $E$(SD). The summed SD strength $S^-$(SD) is derived from the measured  ($^3$He,$t$)  CER spectra at $\theta\approx$ 2 deg. where the $l$=1 cross section gets maximum. The strengths, including the QFS with $\delta l=\delta n=1$ for the DBD nuclei are around 50$\pm{7}$ $\%$ of the N-based summed strength  of $S^-_{\rm N}$(SD)$\approx \Sigma (2l+1)(N_{\rm n}$/2$\pi$)$R_{\rm n}^2$ with  $N_{\rm n}$ and $R_{\rm n}$ being the effective number and the radius of the neutrons involved in  the 1$\hbar\omega (\delta l$=1) neutron to proton transitions, as shown in Fig. 3.  The summed strengths measured by the (p,n) reactions for medium-heavy nuclei    
are about a half of the Tamm Donkoff limit and the N-based QRPA (QRPA(N)) calculations \cite{gaa81}. These data indicate the quenching effect due to the non-nucleonic $\Delta$ effect.

SD and $\delta l$=1 NMEs for low-lying states in medium heavy nuclei are smaller than nuclear model NMEs with NN correlations, suggesting the possible $\Delta$ effects due to non-nucleonic N$\Delta$ correlations \cite{eji82,eji83}. The SD NMEs are also discussed in terms of the shell model, and are shown to be quenched by a coefficient around 0.77 \cite {bro88}. 

In short,  the summed GT and SD strengths are reduced from the sum rule limits by a  factor around $(K_{\Delta})^2$ with $K_{\rm \Delta}$=0.7$\pm{0.07}$ being the quenching coefficient.  Here the error includes the systematic ones in subtracting the small non-GT components with $\delta l \ne 0, \delta n \ne 0$, and also the uncertainty in getting the NMEs from the CER cross section. The tensor component with $<$[Y$_2\sigma]_1>$ is not more than a few $\%$.   
Noting that the sum rule is independent of nuclear models with  nucleonic (NN) correlations, the deviation from the rule indicates some non-nucleonic correlation, which is mainly the N$\Delta$ resonance effect as shown later. The GT sum rule in the nucleon region is about 50 $\%$ smaller than the limit of 3($N-Z$) \cite{ike63} because of the 50 $\%$ shifted to the non-nucleonic GR$_{\rm \Delta}$ region.

\begin{figure}[t]
\hspace{-2.3cm}
\vspace{-0cm}
\includegraphics[width=0.62\textwidth]{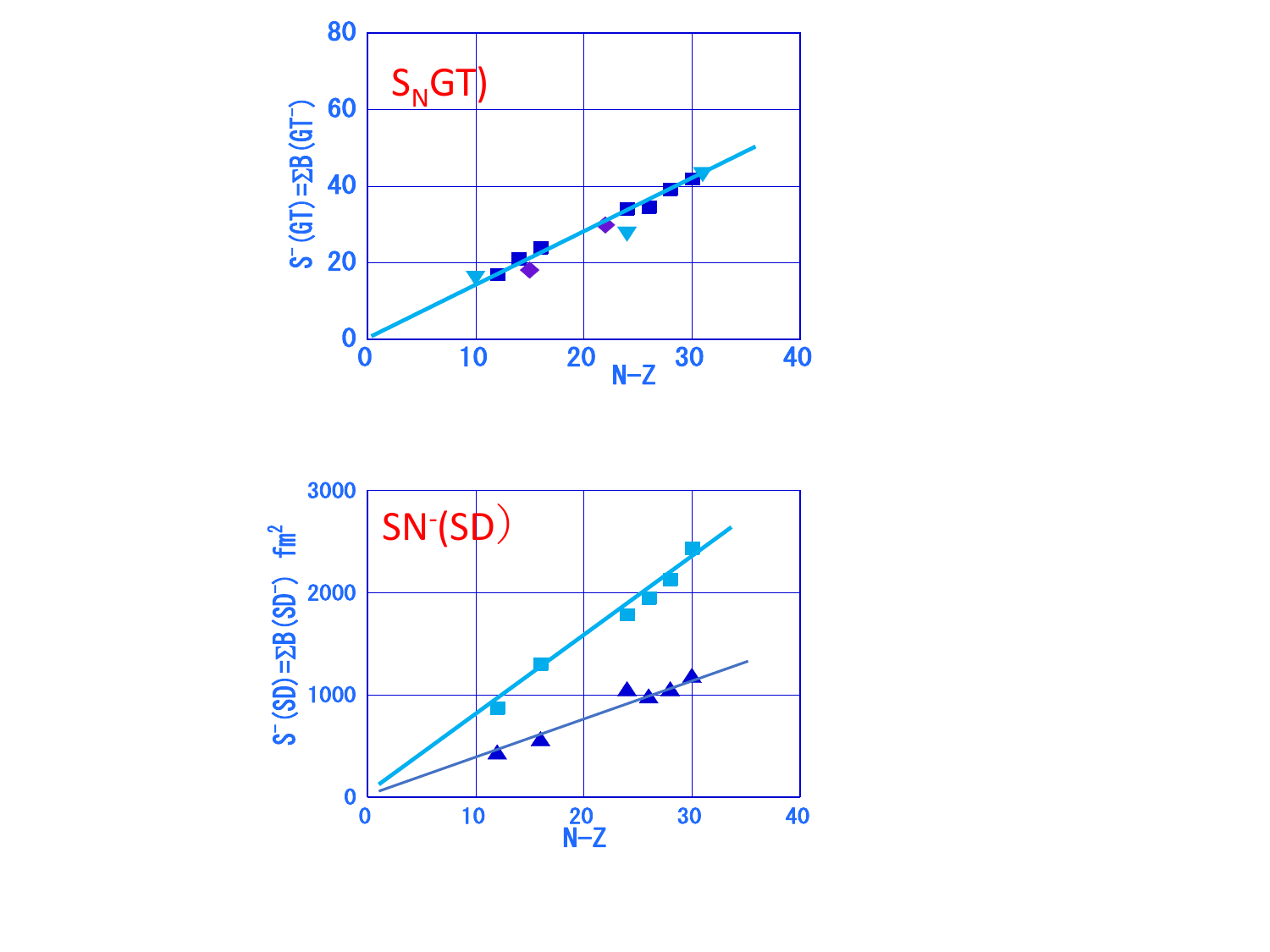}
\hspace{-0cm}
\vspace{-1.3cm}
\caption{Top: Summed GT strength of $S^-$(GT). Blue squares: ($^3$He,$t$) on DBD nuclei. Blue diamonds: ($^3$He,$t$) on Sn isotopes. Light blue squares: (p,n). 
Solid thin line:  $S_{\rm N}$(GT)=3$(N-Z)$. Thick line: 0.47 $S_{\rm N}$(GT).
Bottom: Summed SD strength of $S^-$(SD). Blue triangles: ($^3$He,$t$) on DBD nuclei. Light blue squares and thick line: sum rule of $S^-_{\rm N}$(SD). Thin line: 0.50 $S^-_{\rm N}$(SD).
\label{figure:fig3}} 
\vspace{-0.6cm}
\end{figure}

\section {Non-nucleonic effects and QRPA(N$\Delta$)}

\subsection {N$\Delta$ coupling and $\tau\sigma$ quenching} 

 Experimental $\tau\sigma$ NMEs involved in axial-vector $\beta$ and EM transitions are known to be smaller than NMEs based on thoeretical models with various nucleonic (NN) correlations, suggesting appreciable non-nucleonic effects. 
Non-nucleonic effects have been extensively discussed theoretically as stated in section I. Some of them are made in terms of the $\Delta$N coupling (core-polarization) effect in \cite{boh81,eji82,tow87,ost92,kir99,cat02} and others recently in terms of the 2B current effect using the very advanced effective field theory in \cite{gys19,men11,cor24}. The non-nucleonic $\Delta$ effect on axial-vector weak transitions is discussed in terms of the effective weak coupling of $g_{\rm A}^{eff}$  in \cite {eji19} and refs therein, and also in \cite{eji14,eji15}.

The $\Delta$N coupling theory relies on effective N$\Delta$ interaction, and explains more or less the the quenched $\tau\sigma$ NMEs as observed.  
The 2B (meson-exchange) quenching effects are non-nucleonic effects associated with excited nucleons (i.e. non-nucleons as $\Delta$ and other baryons)  that couple with nucleons via the meson-exchange  interactions. In the present case of the $\tau\sigma$ NMEs, $\Delta$ is the
 only one non-nucleon that couples strongly with the nucleon via the $\pi$-exchange interaction.  Thus, the origin and the effect of the 2B quenching are nealy the same as those of the $\Delta$N quenching, although the theoretical ways of evaluating them are different. These thoretical approaches lead to the severe quenching for the $\tau\sigma$ NMEs and also for the summed $\tau\sigma$ strengths in the N region. 

On the other hand, the present quenching effect is based on the experimental reduction (quenching) of the summed $\tau\sigma$ strengths in the N region. Since the $\Delta$ involved in the $\Delta$N interaction and in the 2B current is located far (300 MeV) above the N region, the quenchin effects for the NMEs in low-lying ( a few MeV) state and for the summed strength in the N region are almost same. Accordingly the present experimental quenching coefficient and the theoretical N$\Delta$/2B quenching coefficient are considered to agree with each other within the experimental and theoretical uncertainties.  
 
In fact, the 2B effects for $^{100}$Sn and for the summed GT strength in the  
$A$=90 nucleus are similar to the present effect of 0.7$\pm{0.07}$ for the similar mass medium heavy nuclei. It is interesting to see how the 2B calculations for DBD nuclei reproduce the present experimental $\tau\sigma$ strengths.

The 2p-2h correlations \cite{ber82,gam20} are the nucleonic NN correlations, which are not included in simple QP and QRPA models with 1p-1h correlations, and modify the $\tau\sigma $ NMEs for low-lying states.  They spread and shift some $\tau\sigma$ strengths from the  GR$_{\rm N}$ region around 10-15 MeV to the higher excitation region above 15 MeV within the N region. Thus it is different from the present non-nucleonic $\Delta$ and 2B  correlations that shift the $\tau\sigma$ strengths to the non-nucleonic $\Delta$ region at 300 MeV far beyond the N region.  Thus the 2p-2h correlation does not quench the summed $\tau\sigma$ strength. 

Then, we discuss how the non-nucleonic effect shown in the summed $\tau\sigma$ strengths may be well explained by using the GR$_{\Delta}$ as explained in section I. The strong repulsive N$\Delta$ interaction gives rise to the giant resonance of  GR$_{\Delta}$, which is a coherent sum of the 2$A$ N$\Delta$ amplitudes with $A$ being the mass number of the nucleus, and the $\Delta$-origin quenching effect  is exclusively incorporated in the GR$_{\Delta}$ effect. The  merits of using the GR$_{\Delta}$ concept is to help understand the effect as given in Fig.1 and in the following subsection and to be linked to the experimental GR$_{\rm N}$ and GR$_{\Delta}$.

So far, most QRPA(N), ISM, IBM, and other model calculations for DBD NMEs take explicitly realistic  nucleonic NN  $\tau\sigma$ correlations, but not explicitly non-nucleonic N$\Delta$ correlations \cite{eji19}. QRPA(N) takes into accounts so many relevant NN correlations as extensively used for $\beta$, DBD and others \cite {eji78,eji00,eji19}. In these QRPA(N), the non-nucleonic N$\Delta$ interaction is not explicitly included in the model, and then the $\tau\sigma$ NN interactions are adjusted so as to reproduce the observed GR$_{\rm N}$ energies, and the calculated summed GT strength is not quenched because the N$\Delta$ interaction is not explicitly included.

\subsection {NN-N$\Delta$ QRPA for GR$_{\Delta}$}

 Let us use a schematic QRPA(N$\Delta$) model with both the NN and  N$\Delta$ $\tau\sigma$ interactions to evaluate gross effects of their correlations on the summed GT strengths and the $\beta$ and $\beta\beta$ NMEs in the N region of $E\leq30$ MeV.   In fact, the N$\Delta$ effects have been discussed also in the QRPA calculations \cite{gaa81,ost92,kir99}. 
The interaction is expressed as 
\begin{eqnarray}
V=g'_{NN}C V_{12}\sigma_1\sigma_2 \tau_1\tau_2  
+ g'_{N\Delta}C' V_{12}
 S_1\sigma_2T_1\tau_2, 
\end{eqnarray}
where $g'_{NN}$ and $g'_{N\Delta}$ are the NN and N$\Delta$ interaction coefficients,  $V_{12}$=$\delta^3(r_1,r_2)$,  $C$=392 MeV fm$^3$ and $C'$=$(f_{\pi N\Delta}/f_{\pi NN})C$=$2C$, and   $T$ and $S$ are for the $\Delta$ isospin and spin \cite {ost92,kir99,bro81}. In the present case of the $\tau ^-$ transition, the particle hole excitations involved are
 n to p, n to $\Delta^+$ and p to $\Delta ^{++}$ for the forward correlations and p to n, p to $\Delta ^0$ and n to $\Delta^-$ for the backward ones.  

In the present DBD and other medium-heavy nuclei with the large neutron excess, the valence neutron shell is so separated from the proton one that the cross term ($D^2$ in \cite{eji68}) is around or less than 0.03. So we assume $D^2$=0.  Then,  using $\chi_N=48~g'_{NN'}$ MeV and $\chi_{\Delta}=96~g'_{N\Delta}$ MeV for the DBD nuclei with the density of $\rho_0$=1.21 fm$^{-3}$, the dispersion equation for the present medium-heavy DBD nuclei with the large $N-Z$  is given by 
\begin{eqnarray}
\frac{\chi_N}{A}
\Sigma_i \frac{|<\phi^-_i||\sigma\tau^-||0>|^2}{\epsilon_i-\epsilon}+
\frac{\chi_\Delta}{A} \Sigma_j\frac{|<\psi^-_j||ST^-||0>|^2}{\epsilon_{\Delta j}-\epsilon}
\nonumber 
\end{eqnarray}
\begin{eqnarray}
+\frac{\chi_\Delta}{A} \Sigma_k \frac{|<\psi^+_k ||ST^+||0>|^2}{\epsilon_{\Delta_K}+\epsilon}
= -1,
\end{eqnarray}
where $\phi^-_i$ stands for the  n$_i^{-1}$p$_i$  state, 
 $\psi_j^-$ for the n$_j^{-1}\Delta_j^+$ and p$_j^{-1} \Delta_{j}^{++}$ states
 and $\psi^+$ for the n$^{-1}_k\Delta_k^-$ and p$^{-1}_k\Delta_{k}^0$ states.
 $\epsilon_i, \epsilon_j$ and $\epsilon_k$ are their energies and $\epsilon$ is the eigen energy. The backward p$^{-1}$n correlation is assumed to be blocked by the large n excess in the DBD nuclei. Contributions from the $\Delta$ at around 300 MeV are given by the second and third terms of eq. (3).  Their sum is  given by
\begin{eqnarray}
  \kappa_{\Delta} \approx \frac{\chi_{\Delta}}{A} [\frac{2(A+0.33A)}{300 \rm MeV}]=0.009\chi_{\Delta}/{\rm MeV},
\end{eqnarray}
where $\kappa_{\Delta}$ stands for the $\Delta$  $\tau\sigma$ susceptibility. Then using $\kappa_{\Delta}$, the dispersion equation is rewritten as 
 \begin{eqnarray}
\frac{\chi_N}{A(1+\kappa_{\Delta})}
[\Sigma_i \frac{|<\phi_i||\sigma\tau^-||0>|^2}{\epsilon_i-\epsilon}]= -1,
\end{eqnarray}
where $\chi_N$/$(1+\kappa_{\Delta}$) is the renormalized NN interaction that includes the $\Delta$ isobar effect. The summed GT strength in the N region is quenched by the same coefficient of
\begin{eqnarray}
K_{\Delta}=1/(1+\kappa_{\Delta}). 
\end{eqnarray}

The quenching coefficient of $K_{\Delta}$=0.7$\pm0.07$ derived experimentally from the summed GT and  SD strengths corresponds to the suaceptibility of $\kappa_{\Delta}\approx$0.43 in eq. (6). This is just expected from $g'_{\rm N\Delta}\approx$0.5 as in the J\"{u}rich-Tokyo potential, \cite{ost92} and the N$\Delta$ interaction $\chi_{\Delta}$=48 MeV in eq. (4).

The quenching due to the GR$_{\Delta}$ effect is a kind of the $\tau\sigma$-type N$\Delta$ core polarization with $\kappa_{\Delta}$ being the $\tau\sigma$-N$\Delta$ susceptibility (polarizability) \cite{eji82,tow87,ris89}.  The present susceptibility is nearly the same as the one derived theoretically in \cite {ris89}. 

\begin{figure}[ht]
\vspace{-0cm}
\includegraphics[width=0.7\textwidth]{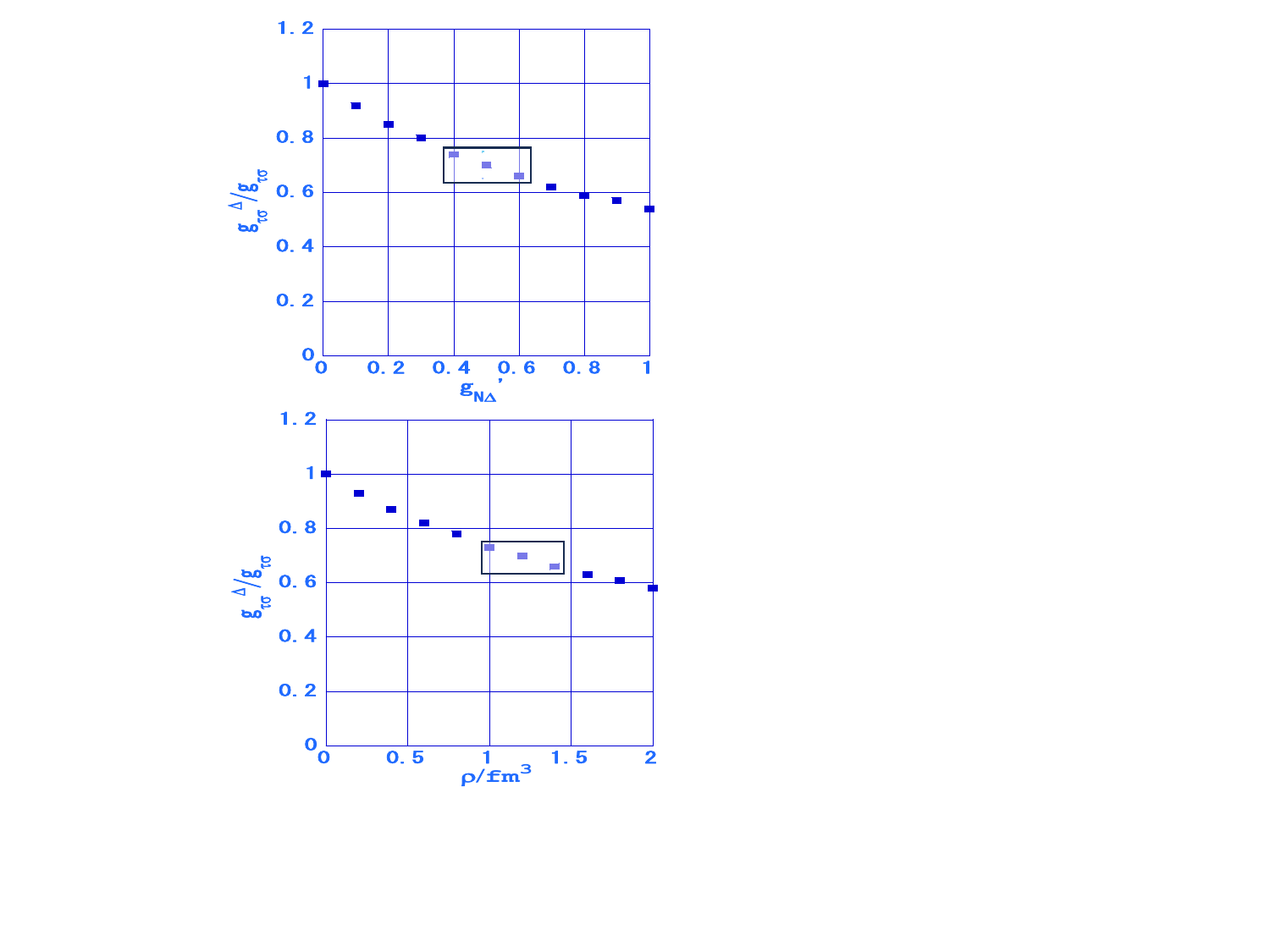}
\vspace{-2.0cm}
\caption{Top: Axial-vector ($\tau\sigma$) effective coupling against the N$\Delta$ interaction parameter $g_{\Delta}'$ at the density $\rho$=1.21 fm$^{-3}$. Bottom: Axial-vector ($\tau\sigma$) effective coupling against the density $\rho$ with the N$\Delta$ interaction parameter $g_{\Delta}'$=0.5. The rectangles correspond to the observed regions of  $g_{\tau\sigma}^{\Delta}$/$g_{\tau\sigma}$=0.7$\pm$0.07.
\label{figure:fig4aa}} 
\vspace{-0.3cm}
\end{figure}
The non-nucleonic reduction effect of GR$_{\Delta}$, which is far above the N region, is considered to be common for all nuclear $\tau\sigma$ components of the weak, electromagnetic and nuclear interaction NMEs in the N region. It is expressed by using the reduced coupling of $g^{\Delta}_{\tau\sigma}$  as 
\begin{eqnarray}
g_{\tau\sigma}^{\Delta}\approx g_{\tau\sigma}(1+\kappa_{\Delta})^{-1}\approx0.7g_{\tau\sigma} 
\end{eqnarray}
where
$g_{\tau\sigma}$ is the coupling for a free nucleon. In case of the weak $\tau\sigma$ NME, the 
$g_{\tau\sigma}$ is the axial-vector weak coupling of $g_{\rm A}$ for a free nucleon. It is given as $g_{\rm A}=1.27 g_{\rm V}$ with $g_{\rm V}$ being the vector coupling. 
 
The $\Delta$ mixing amplitude $a_i$ is of the order of $(\chi_{\Delta}/A)$/300 MeV$\approx$ 1.5 10$^{-3}$ per nucleon, and the $\Delta$ probability is as small as $a_i^2\approx2\times10^{-6}$, but the coherent sum of $a_i$ over 2$A\approx$200 of the $\Delta$s excited from $A\approx$100 of Ns (nucleons) is of the order of $\kappa_{\Delta}\approx$0.3, resulting in the severe quenching coefficient around 0.7 as observed.   

Since the N$\Delta$ interaction depends on the interaction parameter $g_{\rm N\Delta}'$ and  the density $\rho$, $g_{\tau\sigma}^{\Delta}$/$g_{\tau\sigma}$ depends on  $g_{\Delta}'$ and $\rho$ as shown in Fig.4. The observed quenching coefficient is just as expected from the appropriate N$\Delta$ coupling  around $g_{\rm N\Delta}'\approx$0.5 and the known nuclear density  around $\rho\approx$1.25 fm$^{-3}$. The similar $\rho$ dependence is seen in the analysis in terms of the 2B effect \cite{men11}. 

 The present schematic QRPA(N$\Delta$) analysis with both NN and N$\Delta$ interactions is limited on the gross effect of the GR$_{\Delta}$ and GR$_{\rm N}$. Since $\Delta$ is isolated from the N region and is only one resonance that couples strongly with N via the $\tau\sigma$ interaction, the
 non-nucleonic quenching effect on the $\tau\sigma$ NME is mostly ($\ge 90\%$) taken into account by $K_{\Delta}$  and $g_{\tau\sigma}^{\Delta}$ in eqs (6) and (7). 
 
\subsection {NN and N$\Delta$ GRs} 

The N$\Delta$ interaction  pushes down the GR$_{\rm N}$ in energy, while the NN one  pushes up that. The GR$_{\rm N}$(GT) energy is calculated by using eq.(3) with the N$\Delta$ interaction of $g'_{N\Delta}=$0.5 and the NN interaction of $g'_{\rm NN}$=0.62 ($\chi_{\rm N}\approx$30 MeV) as in the J\"{u}lich Tokyo potential and others \cite{ost92,kir99}. The calculated values reproduce well the observed GR$_{\rm N}$(GT) energies as shown in Fig. 5. They are given as $E_{\rm CA}$(GT)$\approx$0.22($N-Z$)+9.3 MeV.  The NN interaction that would fit the observed GR$_{\rm N}$(GT) energy without the N$\Delta$ interaction would be around $\chi'_{\rm N}\approx$21 MeV, which is just $K_{\Delta} \times \chi_{\rm N}\approx$0.7 $\times$30 MeV with the N$\Delta$ interaction.   

\begin{figure}[ht]
\hspace{0.5cm}
\vspace{-2.4cm}
\includegraphics[width=0.6\textwidth]{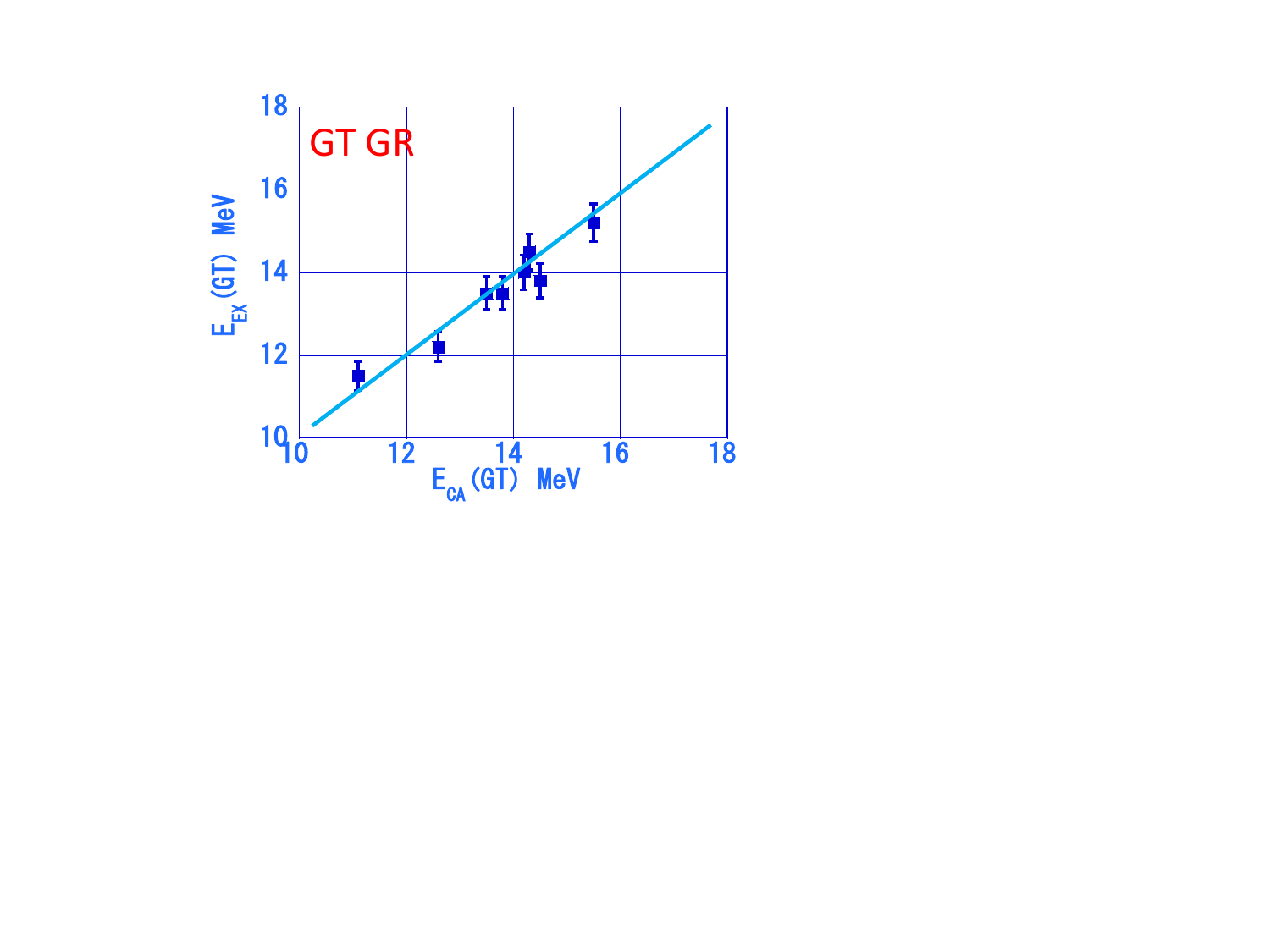}
\vspace{-2.0cm}
\caption{Experimental GR$_{\rm N}$(GT) energies  against calculated ones.  Line $E_{\rm EX}$(GT) = $E_{\rm CA}$(GT). }
\vspace{-0.0cm}
\end{figure}

The SD mode is the $\tau\sigma$ excitation over 1 $\hbar\omega$. The GR$_{\rm N}$(SD) energies for DBD nuclei are extracted from the CERs data \cite{aki97,thi12,thi12b,pup11,gue11,thi12a,pup12}. The GR$_{\rm N}$(SD) is at about 1$\hbar\omega$ above the GR$_{\rm N}$(GT), and the SD QP NMEs  follow the similar reduction (quenching) as the GT ones \cite{eji22}.

The GR$_{\Delta}$ is pushed up in energy due to the repulsive N$\Delta$ interaction from the GR$_{\rm N}$. The GR$_{\Delta}$ energy is expressed by using the unperturbed  energy $E_{\Delta}$(0) for $\Delta$ in the nucleus and the N$\Delta$ interaction energy of $\chi_{\Delta}$,
\begin{eqnarray}
E({\rm GR}_{\Delta})=E_{\Delta}(0)+\frac{\chi_{\Delta}}{A}\frac{4Z+2A}{3},
\end{eqnarray}
Using  $(4Z+2A)/3A\approx1.2$ for the DBD nuclei, one gets $E$(${\rm GR}_{\Delta})\approx$$E_{\Delta}$(0)+58 MeV.  Assuming the $\Delta$ binding energy of 20 MeV as the nucleon one, the unperturbed 
$\Delta$ energy is evaluated as $E_{\Delta}$(0)$\approx$276 MeV, and the $\Delta$ GR energy as $E$(GR$_{\Delta})\approx $334 MeV.  The GR$_{\Delta}$ energy  has been studied by using photo nuclear reaction \cite{cat02,ahr88}. The average excitation energy for heavy nuclei with the similar $(4Z+2A)/3A\approx1.2$ as for DBD nuclei is around 330-340 MeV, although the resonance energy is not well defined due to the large intrinsic and spreading widths of $\Delta$. The cross sections are proportional to $A$. These are consistent with the present GR$_{\Delta}$ with the $\chi_{\Delta}$.

  \section{Quenching of $\tau\sigma$ NMEs for low-lying states}.

The  GT and SD NMEs for the low-lying QP states are reduced due to the NN and N$\Delta$ $\tau\sigma$ correlations. The GT strength is partly shifted from the QP to the GR$_{\rm N}$(GT) around 11-15 MeV  and partly to the GR$_{\Delta}$(GT) above 300 MeV. Then, the NME $M$ for the  ground state GT transition is given as \cite{eji68,eji78,eji00}
\begin{eqnarray}
M=K_{\rm N\Delta}M_{\rm QP}, ~~~K_{\rm N\Delta}=\frac{1}{1+\kappa_{\rm N} + \kappa_{\Delta}},
\end{eqnarray}
where $\kappa_{\Delta}\approx$0.43 as discusses in section 3 and $\kappa_{\rm N}$ is the $\tau\sigma$-NN susceptibility due to the GR$_{\rm N}$ as discussed extensively in \cite{eji78,eji00}.   In the medium-heavy nuclei $\kappa_{\rm N}$ is around 2 \cite{eji78}, but it depends much on the
 nuclear structure. Then, assuming $\kappa_{\rm N}\approx$2, the reduction coefficient is $K_{\rm N\Delta}$=1/(1+2+0.43)$\approx$ 0.3. The $\tau\sigma$ NME for the ground QP state is reduced by a coefficient  around 0.3 with respect to the simple QP NME due to the distractive couplings of GR$_{\rm N}$ and GR$_{\Delta}$. 
 We get the same quenching coefficient by using 
\begin{eqnarray}
K_{\rm N\Delta}=K'_{\rm N}\times K_{\Delta},
\end{eqnarray}
where $K'_{\rm N}=1/(1+\kappa_{\rm N}')$ and $\kappa_{\rm N}'=\kappa_{\rm N}K_{\Delta}\approx 0.7\kappa_{\rm N}$.\\

 We have discussed the non-nucleonic quenching effect of the GR$_{\Delta}$ on the basis of the experimental summed strengths. Actually, the non-nucleonic effect has been discussed theoretically since 1970 mainly in terms of the $\Delta$ effects \cite{ose79,boh81,eji82, sag82,ost85,tow87,ost92,kir99,cat02}, and also the 2B and the exchange-current ($\pi$-exchange between 2B) effects \cite{tow87,ris89, gys19,men11,cor24}. These are mostly on weak GT NMEs. The $\Delta$ effect on the weak SD NMEs is discussed in \cite{eji82}.

Now we discuss briefly effects of the GR$_{\Delta}$ and the GR$_{\rm N}$  on
 $\beta$ and $\beta\beta$ NMEs for low-lying QP states in DBD nuclei.  Experimental GT and SD NMEs for the ground state $\beta$/EC transitions in
 medium-heavy  DBD and other nuclei are smaller by a reduction coefficient $K_{\rm EX}\approx 0.21\pm{0.03}$ with respect to the simple QP NMEs without NN and N$\Delta$ correlations
 \cite{eji15,eji14}. 

The reduction coefficient $K_{\rm N}'\approx 0.4$  is found to be due to the NN 
$\tau\sigma$ (GR$_{\rm N}$)
 and other NN correlations in the QRPA(N) with the realistic G-matrix NN interactions.  \cite{eji15,eji14}. Here the particle-hole interaction parameter of $g_{\rm ph}$ is adjusted so as to reproduce the experimental GR$_{\rm N}$ energies. The coefficient of
 $K_{\rm EX}\approx$0.21 is smaller than the product of $K_{\rm N}'\approx$ 0.4 and $K_{\Delta}\approx$ 0.7 (see eq. (10)),
 suggesting further reduction around 0.8 due to such nucleonic and non-nucleonic effects that are not well included in that QRPA(N) model  \cite{eji15,eji14} for the nucleonic correlation and the present $K_{\Delta}$ for the non-nucleonic correlation. Since most of the non-nucleonic effect is included well in the present $K_{\Delta}$ as explained in sec. III,  this reduction is likely due to such nucleonic effects 
as the 2p-2h correlation \cite{ber82,gam20} and  other NN correlations  that are not included in that QRPA(N) model.  These effects depend much on the individual QP ground state, being small in case of the ground state isolated in energy from other NN states.

The axial-vector $\beta \beta$ NME is expressed as $M^{0\nu}_{\rm A}$= ($g_{\rm A}^{eff})^2M^0_{\rm A}$, where $M^0_{\rm A}$ is
 the model NME.  Here the effective coupling of  $g_{\rm A}^{eff}$ in units of $g_{\rm A}$ for a free nucleon is introduced to incorporate effects which are not included in the model \cite{eji22}. 
 Using $M^0_{\rm A}$ and $M^0_{\rm F}$ derived by the QRPA (N) \cite{eji22} and the value of
 0.8 $K_{\Delta}$ =0.8$\times$0.7 for $g_{\rm A}^{eff}$, 
 we get $ M^{0\nu}\approx5.2 - 0.025A$,
 with $A$ being the mass number. This is close to the NMEs in \cite{eji22} with similar $g_{\rm A}^{eff}$, and those with the 2BC  \cite{men11} .
 
In any way it is indispensable for reliable $\beta$/EC and DBD NME calculations for individual medium heavy nuclei to include exactly all relevant 1p-1h, 2p-2h, and other nucleonic correlations as well as the non-nucleonic GR$_{\Delta}$ effects  that affect the axial-vector, vector and tensor DBD NMEs. This is beyond the scope of the present paper, which discusses mainly the non-nucleonic GR$_{\Delta}$ effect common to all medium heavy nuclei.

\section{Discussions and Concluding Remarks}

GT and SD $\tau\sigma$ NMEs for the medium-heavy DBD nuclei are investigated on the basis of the  experimental $\tau\sigma$ summed strengths measured by the medium energy CERs.
Summed GT and SD strengths in the N region  are shown to be  quenched  by  a factor around $(K_{\Delta}=0.7\pm{0.07})^2$  with respect to the sum rule limits for nuclei composed by nucleons  without non-nucleonic  N$\Delta$ correlations, i.e. the sum rule \cite{ike63} in case of GT.   

The quenching of the $\tau\sigma$ NME in the N region is based on the reduction of the summed strengths in the nucleonic (N) region and thus is due to the non-nucleonic $\Delta$ effect in the present case of the $\tau\sigma$ NME.  The measured quenching is shown by using a schematic QRPA(N$\Delta$) to be explained by the destructive coupling (interference) with the GR$_{\Delta}$. This is a kind of the $\Delta$ polarization effect and the main part of the 2B/exchange current effect discussed in other nuclei. The $\Delta$ effect is mainly represented by the GR$_{\Delta}$ effect.

The present GR$_{\Delta}$  explains well how the small $\Delta$ components in the GR$_{\Delta}$ lying far in energy above the N region reduces the $\tau\sigma$ strength in the N region by acting coherently. The GR$_{\Delta}$ is shown to be consistent with the experimental GR$_{\Delta}$ and GR$_{\rm N}$ energies. 

The present quenching coefficient is based on the missing summed $\tau\sigma$ strengths by the CERs. It could be around 0.77 if some 10 $\%$ of the sum rule \cite{ike63} would be located beyond the present measurement for the N region. On the other hand, the present GR$_{\Delta}$ and other $\Delta$ and 2B machanisms, which are based mainly on the $\Delta$ with the strong non-nucleonic N$\Delta$ coupling, would not be appropriate if the summed GT strength in the N region would agree exactly with the sum rule \cite{ike63,wak97}.

The quenching coefficient derived from the experimental summed $\tau\sigma$ strengths, together with the experimental GR$_{\rm N}$ and GR$_{\Delta}$ energies, are consistent with the QRPA(N$\Delta$) evaluations using the  N$\Delta$  and  NN interactions \cite{ost92}. It is noted here that the GR$_{\Delta}$ is only the axial-vector ($\tau\sigma$) non-nucleonic resonance that couples strongly  with the axial-vector NMEs in the N region. 

The quenching effects on the axial-vector $\beta$ and $\beta\beta$ NMEs are expressed as
$M(\beta)$=($g_{\rm A}^{\Delta}/g_{\rm A})M_{\rm N}(\beta)$ and 
$M(\beta\beta)=(g_{\rm A}^{\Delta}/
g_{\rm A})^2M_{\rm N}(\beta\beta)$ with $g_{\rm A}^{\Delta}/g_{\rm A}$=$K_{\Delta}$=0.7$\pm{0.07}$,
 and $M_{\rm N}(\beta)$ and $M_{\rm N}(\beta\beta)$ are the axial-vector $\beta$ and $\beta\beta$ NMEs with all relevant NN correlations. The quenching effect due to the GR$_{\Delta}$, which is located far beyond the nucleon region of $E$= 0-30 MeV, is common to all $\tau\sigma$ NMEs in the medium-heavy nuclei, being not dependend on individual nuclear structures.

The GR$_{\Delta}$  effect on axial-vector NMEs for the medium-heavy DBD nuclei with $A$=76-136 is discussed in the present work. The interactions used are $\chi'_{\rm N}$=21 MeV (=0.7$\chi_{\rm N}$) and $\chi_{\Delta}$=48 MeV for all nuclei, and thus the quenching coefficient is 0.7 for all the medium heavy DBD nuclei. The quenching effect gets much less at light nuclei in case of the $A$ dependent  interaction proportional to A$^{0.3}$ as suggested in \cite{hom96}.  In this case the quenching coefficients  for light nuclei are $g_{\tau\sigma}^{\Delta}$/$g_{\tau\sigma}\approx$0,9, 0.82, 0.79, and 0.77 for nuclei with $A$=5, 10, 20, and 30. Actually, the quenching coefficients evaluated for light nuclei with $A$=15-38 are 0.9 - 0.7 \cite{ose79}, and a similar feature is seen in \cite {gys19}. The coefficients for the medium-heavy DBD nuclei with $A$=76-136 are 0.71-0.68, being nearly the same as 0.7 for the present constant interaction.  The $A$ dependence may reflect the density dependence in Fig. 4. So interesting is to measure the summed GT strength in light nuclei to see the $A$ dependence of the GR$_{\Delta}$ effect.

The present work is mainly on the axial-vector ($\tau\sigma$) component in $\beta$ and $\beta\beta$ NMEs. In fact, $\beta$ and $\beta\beta$ NMEs include the vector ($\tau$) component, which is considered to be not much quenched because of no coupling with GR$_{\Delta}$.  It is important to measure them experimentally to see if any quenching in there and to validate the theoretical model calculation for them. Gamma rays from isobaric analogue states excited by CERs are used to study vector NMEs \cite{eji23}. OMC (ordinary muon capture), which is a kind of the lepton CER of ($\mu,\nu_{\mu}$), is useful to study both the vector and axial-vector NMEs up to around 50 MeV \cite{eji19,has18}. 

The present GR$_{\Delta}$ effect is considered to be the dominant non-nucleonic $\tau\sigma$ correlation. There are so many nucleonic correlations to be exactly taken into the model calculations for accurate evaluations of the $\beta$ and $\beta\beta$ NMEs. Then the CER cross section data are used to check the theoretical models with nucleonic and non nucleonic correlations, as the summed GT and SD strengths are used for the present non-nucleonic GR$_{\Delta}$ effect. 

The GR$_{\Delta}$ efect is associated with the GR$_{\Delta}$ excited by the strong  $\tau\sigma$-type nuclear interaction, and reduce (quench) $\tau\sigma$ components of the weak-, electromagnetic- and strong-interaction NMEs in astro-nuclear physics. Then, the $\tau\sigma$-type NMEs involved in supernova $\nu$-nuclear syntheses, photo-nuclear excitations, isovector spin nuclear reactions and others are  reduced similarly by the quenching coefficient $K_{\Delta}$, which is around 0.7 in medium and heavy nuclei. Thus impact of the GR$_{\Delta}$ on astro-nuclear and particle physics is indeed very large.  It is important to include explicitly and precisely in $\beta\beta$ and other NME calculations the N$\Delta$ interaction in addition to the relevant NN interactions.\\

It is noted that the effect of the present GR$_{\Delta}$ associated with the strong $\tau\sigma$ (isospin spin) coupling is exclusively on the isospin spin components of the weak, electromagnetic and nuclear interaction NMEs.  The $\tau\sigma$ component is the dominant one in cases of the low-energy axial-vector (unique:GT) weak and also in the medium-energy CER NMEs for 1$^+\rightarrow0^+$, and thus they are quenched by the similar coefficient, and the GR$_{\Delta}$ and the 2B effects are nearly the same.  On the other hand, in the other cases as non-unique weak, electro-magnetic, and inelastic nuclear interactions, the vector, the isoscalar, the orbital and even the tensor components 
are involved more or less in addition to the $\tau\sigma$ component \cite{eji00,eji19}. Accordingly, the 2B quenching effects depend much on the relative weights of these components, and  on ndividual nuclear structures, and thus are different from the GR$_{\Delta}$ quenching effects. The 2B effects on electro-magnetic NMEs are discussed in the recent works \cite {miy24,ach24}.

The author thank Prof. T. Sato, Osaka University for valuable discussions.

\end{document}